\documentclass[prl,superscriptaddress,showpacs,amssymb,amsmath,
amsfonts,aps,twocolumn,secnumarabic,floatfix]{revtex4}
\usepackage{graphics}
\usepackage{epsfig}
\begin{document} 
\bibliographystyle{try} 
 
\topmargin 0.1cm 
 
 \title{Pentaquark, cusp and rescattering in single kaon photoproduction off deuterium}

\newcommand*{\JLAB }{ Thomas Jefferson National Accelerator Facility, Newport News, Virginia 23606} 
\affiliation{\JLAB } 

\author{J.M.~Laget}
     \affiliation{\JLAB}

\date{\today} 

\begin{abstract} 
In very well defined part of the phase space, $K N$ and $\Lambda N$ rescattering depend on on-shell elementary matrix elements and on the low momentum components of the deuteron wave function. This provides us with the unique opportunity to study details in the  scattering amplitudes that may have escaped  the analysis of reactions induced on a nucleon target by Kaon and Hyperon beams at low energies. When folded with a typical experimental mass resolution, a narrow state with a width of 1~MeV would contribute by no more than 10\% to the $KN$ mass spectrum.   On the contrary, a cusp would be easily detected near the $\Sigma$ production threshold in the $\Lambda N$ mass spectrum. 
\end{abstract} 
 
 
\maketitle 

An experimental study of exclusive kaon photoproduction off deuterium with high statistics has been recently completed~\cite{g10} at Jefferson Laboratory (JLab). While it was designed
to look for possible pentaquark states, it has opened the way to access the hyperon nucleon interaction in an energy range where the qualities of kaon or hyperon beams are marginal. In very well defined parts of the phase space, the reaction amplitude develops a logarithmic singularity which enhances the cross section. The physical picture is the following. The $\Lambda$, or the $K$, is produced on a nucleon almost at rest, propagates on-shell and rescatters on the second nucleon, also at rest in deuterium, which recoils with the observed momentum $p_n$. Here, the reaction amplitude is on solid ground since it depends only on on-shell elementary amplitudes and on low momentum components of the nuclear wave function~\cite{La81,La05}.  In Ref.~\cite{La05} predictions have been given at high energies and large momentum transfers. This note focuses on kaon production at lower energy. It evaluates the size of a possible pentaquark signal in the $KN$  final state as well as the sensitivity to the $\Lambda N$ amplitude in the vicinity of the  $\Sigma N$ threshold.

\begin{figure}[hbt]
\begin{center}
\hspace{-1.5cm}
\epsfig{file=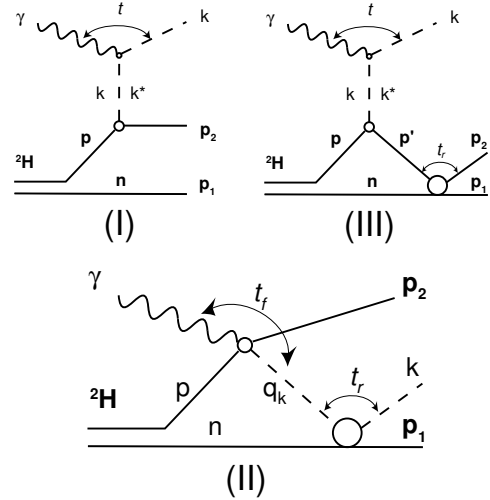,width=3.in}
\caption[]{The relevant graphs in the $\gamma ^2$H$ \rightarrow n K^+ \Lambda$ reaction. I: Quasi-free Kaon production. II: Kaon Nucleon rescattering. III: Lambda Nucleon rescattering.}
\label{graph}
\end{center}
\end{figure}

Fig.~\ref{graph} shows the dominant processes in the photoproduction of kaons off deuterium. We consider as an example the $\gamma ^2$H$ \rightarrow n K^+ \Lambda$ reaction, but the extension to the $\gamma ^2$H$ \rightarrow p K^{\circ} \Lambda$ or $\gamma ^2$H$ \rightarrow n K^+ \Lambda^*(1520)$ channels, as well as $\Sigma$'s production channels, is straightforward.

Graph I represents the quasi-free production of the $K^+$ on the proton, the neutron being spectator. It dominates at low values of the spectator neutron momentum $p_n$, where the cross section takes the simple form~\cite{La05,La81}:
 \begin{eqnarray}
\frac{d\sigma}{d\vec{p_n}d\Omega_{k}}=
(1+\beta_n \cos\theta_n)\rho(|\vec{p_n}|)\frac{d\sigma}{d\Omega_{k}}
(\gamma p\rightarrow \Lambda K^+)
\label{cross_qf}
\end{eqnarray}
where $\beta_n=p_n/E_n$ and $\theta_n$ are the velocity and the angle of the spectator neutron. When the neutron momentum increases, its momentum distribution $\rho(|\vec{p_n}|)$ decreases very quickly and Kaon-nucleon  (graph II) as well as Lambda-nucleon (graph III) rescattering take over above, let's say, $p_n= 300$~MeV/c. 

 The detailed expressions of the corresponding amplitudes are given in ref.~\cite{La05} for pion production at high energy. For the sake of completeness I rewrite them in the kaon production sector and customize the choice of the elementary amplitudes at lower energies.

Let $k=(\nu,\vec{k})$, $p_D=(M_D,\vec{0})$, $p_{k}=(E_{k},\vec{p_{k}})$, $p_1=(E_1,\vec{p_1})$ and $p_2=(E_2,\vec{p_2})$ be the four momenta, in the Lab. system, of respectively the incoming photon, the target deuteron, the outgoing kaon, the slow outgoing neutron and the fast outgoing $\Lambda$. The 5-fold fully differential cross section is related to the square of the coherent sum of the matrix elements as follows:
\begin{eqnarray}
\frac{d\sigma}{d\vec{p_1}[d\Omega_{k}]_{cm2}}&=& \frac{1}{(2\pi)^5}
\frac{|\vec{\mu}_{c.m.}|mm_{\Lambda}}{24|\vec{k}|E_1Q_f}
\sum_{\epsilon,M,m_1,m_2}\left| \sum_{i=I}^{III}
\right.\nonumber \\
&&\left.
{\cal M}_i(\vec{k},\epsilon,M,\vec{p_{k}},\vec{p_1},m_1,\vec{p_2},m_2)
\right|^2
\label{cross}
\end{eqnarray}
where $\epsilon$ is the polarization vector of the photon and where $M$, $m_1$ and $m_2$ are the magnetic quantum numbers of the target deuteron, the outgoing neutron and $\Lambda$ respectively. The norm of the spinors is $\overline{u}u=1$. The amplitudes are computed in the Lab. frame. The cross section is differential in the Lab. three-momentum of neutron, but in the solid angle of the kaon expressed in the c.m. frame of the pair made by the kaon and $\Lambda$. In this frame, the momentum of the kaon is $\vec{\mu}_{c.m.}$ and the total energy is $Q_f= \sqrt{(E_{k}+E_2)^2 -(\vec{p_{k}} +\vec{p_2})^2}$. 

\begin{figure}[h]
\begin{center}
\epsfig{file=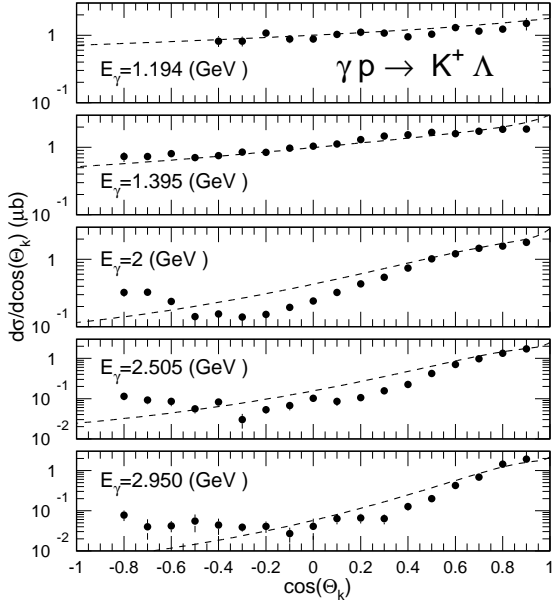,width=3.in}
\caption{The cross section of the elementary reaction $p(\gamma,K^+)\Lambda$. The curve is the prediction of the Regge model~\cite{La97}.  The data has been recently recorded at JLab~\cite{Sch05}.}
\protect\label{nucleon}
\end{center}
\end{figure}

The matrix element of the quasi-free amplitude (graph I in Fig.~\ref{graph}) takes the simple form:
\begin{eqnarray}
{\cal M}_I(\vec{k},\epsilon,M,\vec{p_{k}},\vec{p_1},m_1,\vec{p_2},m_2)=
\nonumber \\
i \sum_{m_p m_l m_s} \sum_{ls} (lm_lsm_s|1M)(\frac{1}{2}m_p\frac{1}{2}m_1|sm_s)
\nonumber \\
u_l(|\vec{p_1}|) Y_l^{m_l}(\vec{p_1})
T_{\gamma p}(\vec{p_2},m_2,-\vec{p_1},m_p)
\label{q_f}
\end{eqnarray}
where $u_0$ and $u_2$ are the $S$ and $D$ components of the deuteron Paris wave function~\cite{PaXX}, and where $T_{\gamma p}$ is the amplitude of the elementary $p(\gamma,K^+)\Lambda$ reaction.  I use the on-shell expression (see Appendix of ref.~\cite{La05}) of the $K$ and $K^*$ Regge exchange amplitudes of ref.~\cite{La97}. In order to improve the agreement with  the differential cross section of the $p(\gamma,K^+)\Lambda$ reaction  recently determined at JLab~\cite{Sch05}, I use a monopole form factor (with the cut-off mass $\Lambda_k=2$~GeV) at the $KN\Lambda$ and $K^*N\Lambda$ hadronic vertices. As shown in Fig.~\ref{nucleon}, this leads to a good description of the differential cross sections, except at very backward angles where $u-$channel exchange mechanisms play a role. The prediction of the original version~\cite{La97} of the Regge model overestimates the JLab data by 20--30~\% and can be found in the experimental paper~\cite{Sch05}.     

The matrix element of the kaon-nucleon rescattering amplitude (graph II in Fig.~\ref{graph}) takes the form:
\begin{eqnarray}
{\cal M}_{II}(\vec{k},\epsilon,M,\vec{p_{k}},\vec{p_1},m_1,\vec{p_2},m_2)=
\nonumber \\
i \sum_{m_n m_p} (\frac{1}{2}m_p\frac{1}{2}m_n|1M)
\int \frac{d^3\vec{n}}{(2\pi)^3} \frac{u_0(n)}{\sqrt{4\pi}}
\frac{1}{q^2_{k}-m^2_{k}+i\epsilon}
\nonumber \\
\frac{m}{E_n}
T_{\gamma N}(\vec{p_2},m_2,-\vec{n},m_p)T_{K N}(\vec{p_1},m_1,\vec{n},m_n)
\nonumber \\
+\mathrm{Deuteron}\; D \; \mathrm{wave}\ \mathrm{part} \;\;
\label{pi_rescat}
\end{eqnarray}

The integral runs on the three momentum of the spectator nucleon in the loop, which has been put on-shell, $n^{\circ}=E_n=\sqrt{\vec{n}^2+ m^2}$, by the integration over its energy $n^{\circ}$. It can be split in two parts:
\begin{eqnarray}
{\cal M}_{II} = {\cal M}_{II}^{on}+{\cal M}_{II}^{off}
\end{eqnarray}

The singular part of the rescattering integral runs between the minimum and maximum values of the momentum of the spectator nucleon in the loop for which the kaon can propagate on-shell:
\begin{eqnarray}
p_{min}(nK)= \frac{P}{Q_s}E_{c.m.}- \frac{E}{Q_s}p_{c.m.} 
\label{pmin_ppi}
\end{eqnarray}
\begin{eqnarray}
p_{max}(nK)= \frac{P}{Q_s}E_{c.m.}+ \frac{E}{Q_s}p_{c.m.} 
\label{pmax_ppi}
\end{eqnarray}
where $E=E_{k}+E_1$, $\vec{P}=\vec{p_{k}}+\vec{p_1}$ and $Q_s=\sqrt{E^2-\vec{P}^2}$ are respectively the energy, the momentum and the mass of the scattering $K^+ n$ pair. The momentum and energy of the outgoing neutron, in the c.m. frame of the $K^+ n$ pair are
\begin{eqnarray}
p_{c.m.}= \frac{\sqrt{(Q^2_s-(m+m_{k})^2)(Q^2_s-(m-m_{k})^2)}}{2 Q_s}
\end{eqnarray}
\begin{eqnarray}
E_{c.m.}= \sqrt{p^2_{c.m.}+m^2} 
= \frac{Q^2_s+m^2-m^2_{k}}{2 Q_s}
\end{eqnarray}
 
It takes the form:
\begin{eqnarray}
{\cal M}_{II}^{on}= \frac{\pi}{(2\pi)^3\sqrt{{4\pi}}} \sum_{m_nm_p}
\frac{1}{2P} (\frac{1}{2}m_p\frac{1}{2}m_n|1M)
\nonumber \\
\int_0^{2\pi} d\phi \int_{|p_{min}(nK)|}^{p_{max}(nK)} nu_0(n) dn
\frac{m}{E_n}
\left[T_{\gamma N}T_{K N}\right]_{q^2_{k}=m^2_{k}}
\nonumber \\
+\mathrm{Deuteron}\; D \; \mathrm{wave}\ \mathrm{part} \;\;
\label{sing_pin}
\end{eqnarray}

The two fold integral depends only on on-shell elementary amplitudes. The weight, $nu_0(n)$, 
selects nucleons almost at rest in the deuterium when the lower bound, $p_{min}(nK)$, of the integral vanishes. This is the origin of the meson-nucleon scattering peak which is therefore on solid grounds. The principal part of the integral vanishes when $p_{min}(nK)=0$ and contributes little to the tail of this peak (see {\it e.g.} ref.~\cite{La05b}).

I take into account both elastic $K^+n$ and charge exchange rescattering:
\begin{eqnarray}
T_{\gamma N}T_{K N}= T_{\gamma p\rightarrow K^+ \Lambda} T_{K^+n\rightarrow K^+n}
-T_{\gamma n\rightarrow K^{\circ} \Lambda} T_{K^{\circ}p\rightarrow K^+n}
\nonumber \\ {}
\label{tg_tk}
\end{eqnarray}
the minus sign coming from the isospin of the deuteron target. Under the assumption that only $K$ and $K^*$ exchange dominate, the $K^+$ and $K^{\circ}$ photoproduction amplitudes are related in the following way
\begin{eqnarray}
\begin{array}{rcrcr}
T_{\gamma p\rightarrow K^+ \Lambda}&=& T^{K}_{\gamma p\rightarrow K^+ \Lambda}
&+& T^{K^*}_{\gamma p\rightarrow K^+ \Lambda} \\
&&&&\\
T_{\gamma n\rightarrow K^{\circ} \Lambda}&=& 0
&-& \frac{g_{\gamma K^*K^{\circ}}}{g_{\gamma K^*K^+}} 
T^{K^*}_{\gamma p\rightarrow K^+ \Lambda}
\end{array}
\label{elem_amp}
\end{eqnarray}
where the minus sign comes from the isospin coefficient at the $K^*N\Lambda$ vertices and where the ratio of radiative constants $g_{\gamma K^*K^{\circ}}/g_{\gamma K^*K^+}=$~1.542 is nothing but the square root of the ratio of the corresponding radiative widths~\cite{PDG}. The isospin decomposition of the $KN$ scattering amplitudes is
\begin{eqnarray}
\begin{array}{rcrcr}
T_{K^+n\rightarrow K^+n}&=& \frac{1}{2}T^1_{K N}&+&\frac{1}{2}T^0_{K N} \\
&&&&\\
T_{K^{\circ}p\rightarrow K^+n}&=& \frac{1}{2}T^1_{K N}&-&\frac{1}{2}T^0_{KN}
\end{array}
\end{eqnarray}

The combination of the elementary matrix elements in eq.~(\ref{pi_rescat}) becomes:
\begin{eqnarray}
\begin{array}{rcr}
T_{\gamma N}T_{K N}&=& T^{K}_{\gamma p\rightarrow K^+ \Lambda}
\left( 0.5 \;T^1_{K N} +  0.5\; T^0_{K N}\right) \\
&&\\
&+& T^{K^*}_{\gamma p\rightarrow K^+ \Lambda}
\left( 1.27 \;T^1_{K N} -  0.271 \;T^0_{K N}\right)
\end{array}
\end{eqnarray}

The isospin $I=0$ and $I=1$ parts of the $K N$ scattering amplitudes can be expressed as:
\begin{eqnarray}
T^I_{K N} = (m_1|f^I(Q_s,t_r) 
+i g^I(Q_s,t_r) \vec{\sigma}\cdot \vec{k}_{\perp}|m_p)
\end{eqnarray}
where $t_r=(p_{k}-q_{k})^2$ is the four momentum transfer at the $K N$ rescattering vertex and $\vec{k}_{\perp}= \vec{p_{k}}\times \vec{q_{k}}$ is the direction perpendicular to the scattering plane. I expand the functions $f^I$ and $g^I$ into S and P partial waves (see for instance ref.~\cite{La93}): I  parameterize them in terms of the experimental phase shifts~\cite{Gi74}, which I used in the analysis of the $pp\rightarrow K^+ p \Lambda$ reaction~\cite{La93,La91}. 

\begin{figure}[hbt]
\begin{center}
\epsfig{file=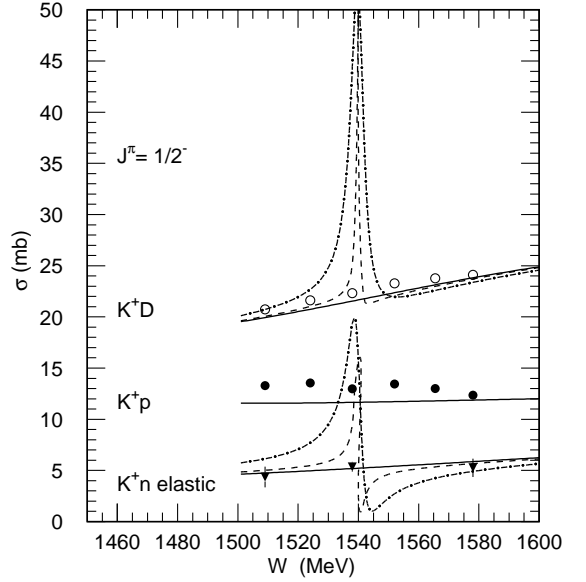,width=3.0in}
\caption[]{The Kaon Nucleon elastic scattering. The full curves correspond to the experimental phase shifts. The dashed and dot-dashed curves include a $J^{\pi}=1/2^-$ resonant state with a mass of 1540~MeV and a width of 1 MeV or 5 MeV respectively. The experimental data come from ref.~\cite{Bo70}.}
\label{Kn_elastic}
\end{center}
\end{figure}

By construction this representation reproduces the  $K^+p$ elastic scattering cross section~\cite{PDG} up to $Q_s \sim$~3 GeV. Above the elastic cross section becomes flat and is better reproduced by the diffractive ansatz which I used in ref~\cite{La05}. Fig.~\ref{Kn_elastic} shows the $KN$ elastic cross sections and the $K^2$H total cross section in the vicinity of a possible pentaquark state~\cite{LEPS,Ste03}. I have added to the non resonant S-wave, isospin $I=0$, experimental amplitude the resonant $J^{\pi}=1/2^-$ amplitude
\begin{eqnarray}
f^0= -\frac{2M_Rg^2_R}{Q^2_s -M^2_R +iM_R\Gamma} \frac{E_n+m}{2m} \;
;\;\;\; g^0=0
\end{eqnarray}
where~\cite{Kw05}
\begin{eqnarray}
\frac{g^2_R}{4\pi}= \frac{M_R\Gamma}{[p_k(E_n+m)]_{c.m.}}
\end{eqnarray}

Similar expressions can be used for a resonant $J^{\pi}=1/2^+$ amplitude, in a P-wave:  
\begin{eqnarray}
f^0= \frac{2M_Rg^2_R}{Q^2_s -M^2_R +iM_R\Gamma}
 \frac{\vec{p_k}\cdot\vec{p_k}}{2m(E_n+m)}
\end{eqnarray}
\begin{eqnarray}
g^0= \frac{2M_Rg^2_R}{Q^2_s -M^2_R +iM_R\Gamma}
 \frac{1}{2m(E_n+m)}
\end{eqnarray}
where~\cite{Kw05}
\begin{eqnarray}
\frac{g^2_R}{4\pi}= \frac{M_R\Gamma}{[p_k(E_n-m)]_{c.m.}}
\end{eqnarray}
In this channel, I have verified that the signal is very similar to the S-wave signal (Fig.~\ref{Kn_elastic}) and to previous predictions~\cite{Ha03} of the isospin $I=0$ cross section. Usually the experimental resolution is broader than the natural width of the resonance. The signal is proportional to the area of the  peak and therefore to the width of the resonance. The analysis~\cite{Ca04,Nu03} of the existing data~\cite{Bo70} within this scheme excludes a resonant state with a width of more that a few MeV. A similar conclusion has been reached in the analysis~\cite{Ar03} of the $KN$ phase shifts.

\begin{figure}[hbt]
\begin{center}
\epsfig{file=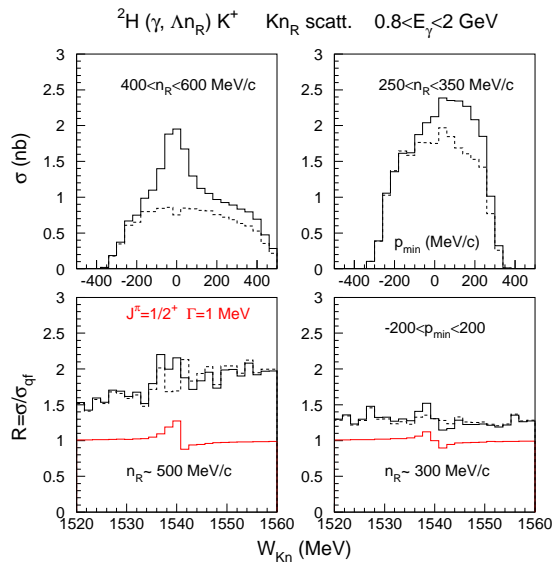,width=3.0in}
\caption[]{Color on line. The $K N$ rescattering sector in the $\gamma ^2$H$\rightarrow K^+ n\Lambda$ reaction. In the two top windows, the dashed and full curves correspond to the quasi free and full model respectively. In the two bottom windows the full and dashed curves represent the ratio of the full model to the quasi free model with and without  a resonant state with a width of $1$~MeV. The flat (red) curve is the ratio between the full and the dashed histograms. See text for the definition of the cuts.}
\label{Kn}
\end{center}
\end{figure}

Figs.~\ref{Kn}  shows various observables which emphasize the kaon nucleon rescattering sector in the $\gamma ^2$H$\rightarrow K^+ n\Lambda$ reaction. The real photon beam end point has been set to $E_{\gamma}=2$~GeV.  The panel shows cross sections integrated over the various bins~\footnote{The events have been binned in the same way as experimental data would be. Each bin mixes events corresponding to different energy in the continuous tagged photon beam. I have chosen to normalize the cross section to one photon in the entire flat photon spectrum. }, within the CEBAF Large Acceptance Spectrometer~\cite{CLAS} (CLAS) fiducial acceptance, by the Monte Carlo procedure that is depicted in ref.~\cite{La05}. Only the geometrical fiducial cut of the Kaon has been taken into account. This is the most important cut since it defines the momentum transfer between the photon and the Kaon. Neither the decay in flight of the Kaon nor the decay $\Lambda \rightarrow p\pi^-$ have been taken into account: they are expected to lead to an almost constant overall reduction factor. The top parts show the distribution of the minimum momentum $p_{min}(nK)$ of the spectator nucleon, in the kaon nucleon scattering loop, for which the kaon can propagate on-shell. On the left, the cut $400<n_R<600$~MeV/c has been applied on the momentum $p_n$ of the neutron in the final state: the kaon nucleon rescattering peak clearly appears at $p_{min}=0$. On the right, the cut $250<n_R<350$~MeV/c has been applied: rescattering effects are small here and the shape of the distribution reflects the kinematics and the detector acceptance. This is a good reference point which emphasizes the quasi-free process.

A further cut $-200<p_{min}(nK)<200$~MeV/c has been applied in the bottom parts of Fig.~\ref{Kn}: it emphasizes kaon nucleon rescattering. The distribution of the mass $W_{Kn}=Q_s$ of the $K^+n$ system is plotted on the left for the high recoil momentum ($\sim 500$~MeV/c) and on the right for the low recoil momentum ($\sim 300$~MeV/c) bands. The dashed curves are the ratio of the full cross section to the quasi-free cross section when no pentaquark contributes. The full curves are the same ratio when a positive parity pentaquark, with $M_R=1540$~MeV and $\Gamma=1$~MeV, contributes. The flat (red) curves in the bottom represent the ratio of the full mass distributions with and without the pentaquark contribution. These ratios minimize the statistical fluctuations, since the Monte Carlo sampling is the same in each histogram. The width of each bin in the $KN$ mass spectrum is $1.6$~MeV, smaller than the CLAS resolution (about 8~MeV). When integrated over $\Delta W_{Kn}=8$~MeV, the signal is 5\% in the high momentum ($n_R\sim 500$~MeV/c) sample and about 0\% in the low momentum ($n_R\sim 300$~MeV/c) sample.

Similar results are obtained when a negative parity pentaquark ($J^{\pi}=1/2^-$) contributes~\cite{La06}: the signal is 15\% in the high momentum ($n_R\sim 500$~MeV/c) sample and 5\% in the low momentum ($n_R\sim 300$~MeV/c) sample.

\begin{figure}[hbt]
\begin{center}
\epsfig{file=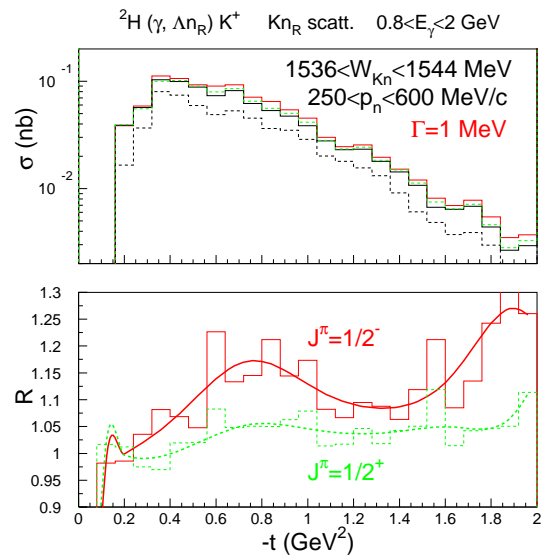,width=3.0in}
\caption[]{Color on line. The angular distribution in the vicinity of a possible pentaquark signal. In the top window, the black dashed and full curves correspond  respectively to the quasi free and full model without pentaquark contribution. The fainted dashed and full curves correspond respectively to the full model with a  positive or a negative parity pentaquark contribution. In the bottom window, the histograms represent the ratio of the full model with and without  a resonant state with a width of $1$~MeV. The curves correspond to a spline fit to each histogram. Full lines: $J^{\pi}=1/2^-$. Dashed lines: $J^{\pi}=1/2^+$. A cut $-200<p_{min}(nK)<200$~MeV/c has been applied.}
\label{Kn_t_dis}
\end{center}
\end{figure}

The top part of Fig.~\ref{Kn_t_dis} displays the angular distribution of the signal in the vicinity of a possible pentaquark signal. The cut at low $-t$ comes the lack of acceptance of CLAS at very forward angles. The effects of the detector acceptance are strongly reduced in the bottom part which shows the ratio of the cross sections with and without the contribution of a possible pentaquark. The signal is larger and increases at large $-t$ (where the cross section is smaller) when $J^{\pi}=1/2^-$. The signal is smaller and flatter when $J^{\pi}=1/2^+$. This is parly due to a different interference pattern.  

This prediction is more stringent than the pentaquark production cross section alone that was evaluated in ref.~\cite{Gu04}. In fact the nonresonant physical background in the vicinity (side bands) of the peak is generated by the same reaction mechanism as the pentaquark itself. Also, the two contributions interfere. The ratio between the area of the resonant peak to the non resonant side bands is a direct measure of the width. This has to be kept in mind when analyzing the data: it is inaccurate to adjust independently the resonant and the non resonant contributions when fitting the data.

$\Lambda N$ rescattering gives also a small flat contribution below the $KN$ rescattering peak. The matrix element of the Lambda-neutron rescattering amplitude (graph III in Fig.~\ref{graph}) takes the form:
\begin{eqnarray}
{\cal M}_{III}(\vec{k},\epsilon,M,\vec{p_{k}},\vec{p_1},m_1,\vec{p_2},m_2)=
\nonumber \\
i \sum_{m_n m_pm'_p} (\frac{1}{2}m_p\frac{1}{2}m_n|1M)
\int \frac{d^3\vec{n}}{(2\pi)^3} \frac{u_0(n)}{\sqrt{4\pi}}
\frac{1}{{p^{\circ}}'-E'_p+i\epsilon}
\nonumber \\
\frac{m}{E_n}
T_{\gamma p}(\vec{p'},m'_{\Lambda},-\vec{n},m_n)
T_{\Lambda n}(\vec{p_2},m_2,\vec{p_1},m_1,\vec{p'},m'_{\Lambda},\vec{n},m_n)
\nonumber \\
+\mathrm{Deuteron}\; D \; \mathrm{wave}\ \mathrm{part} \;\;
\label{p_rescat}
\end{eqnarray}

The integral runs on the three momentum of the spectator neutron in the loop, which has been put on-shell, $n^{\circ}=E_n=\sqrt{\vec{n}^2+ m^2}$, by the integration over its energy $n^{\circ}$. It can be split in two parts:
\begin{eqnarray}
{\cal M}_{III} = {\cal M}_{III}^{on}+{\cal M}_{III}^{off}
\end{eqnarray}

The singular part of the rescattering integral runs between the minimum and maximum values of the momentum of the spectator proton in the loop for which the struck proton can propagate on-shell:
\begin{eqnarray}
p_{min}(\Lambda n)= \frac{P}{W}E_{c.m.}- \frac{E}{W}p_{c.m.} 
\label{pmin_pp}
\end{eqnarray}
\begin{eqnarray}
p_{max}(\Lambda n)= \frac{P}{W}E_{c.m.}+ \frac{E}{W}p_{c.m.} 
\label{pmax_pp}
\end{eqnarray}
where $E=E_2+E_1$, $\vec{P}=\vec{p_2}+\vec{p_1}$ and $W=\sqrt{E^2-\vec{P}^2}$ are respectively the energy, the momentum and the mass of the scattering $\Lambda n$ pair. The momentum and energy of the spectator neutron, in the c.m. frame of the $\Lambda n$ pair are
\begin{eqnarray}
p_{c.m.}= \frac{\sqrt{(W^2-(m+m_{\Lambda})^2)(W^2-(m-m_{\Lambda})^2)}}{2 W}
\end{eqnarray}
\begin{eqnarray}
E_{c.m.}=   \sqrt{p^2_{c.m.}+m^2} 
= \frac{W^2+m^2-m^2_{\Lambda}}{2 W}
\end{eqnarray} 

\begin{figure}[hbt]
\begin{center}
\epsfig{file=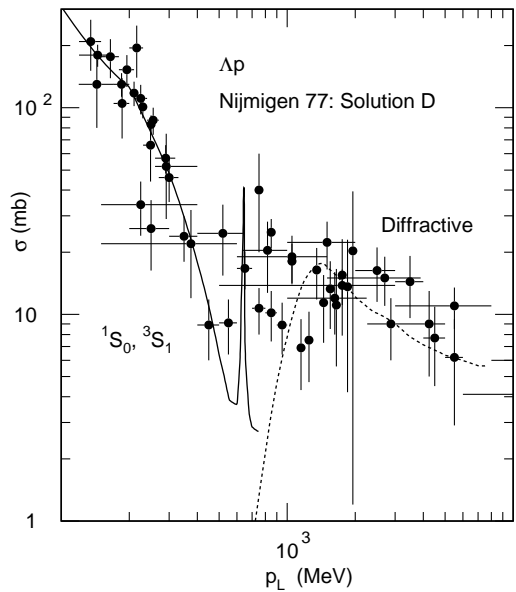,width=3.0in}
\caption[]{The $\Lambda p$ elastic scattering. The full curve corresponds to the low energy phase shift expansion. The dashed curve is the high energy diffractive parameterization.}
\label{Ln_elastic}
\end{center}
\end{figure}

I parameterize the $\Lambda N$ scattering amplitude by its expansion (see {\it e.g.} ref.~\cite{La93}) in terms of the $^1S_0$ and $^3S_1$ waves of the Nijmigen potential~\cite{Na79}. By construction this gives a good description of the available cross sections from threshold up to $p_{\Lambda}\sim$~500 MeV/c (fig.~\ref{Ln_elastic}). Around the $\Sigma$ threshold, $p_{\Lambda}\sim$~650 MeV/c, solution D develops a cusp due to a strong coupling between the $\Lambda N$ and $\Sigma N$ scattering channels. The cusp is less strong or absent in other solutions. In this exploratory study I do not take into account $P$ and $D$ partial waves which may also develop a cusp~\cite{Rij99}: they will be included when the experiment is available. At higher energies, the diffractive ansatz which I used in ref~\cite{La05} is more adequate. The experimental data set~\cite{PDG} is of poor quality. Nijmigen's solution D leads to a good accounting  of Kaon production in $pp$ scattering~\cite{La91,La93} where however it was not possible to disentangle the $\Lambda$ and $\Sigma$ channels. This becomes possible with CLAS which opens up the possibility to map out for the first time the Hyperon Nucleon amplitudes in the vicinity of the $\Sigma$ threshold and higher energy.

\begin{figure}[hbt]
\begin{center}
\epsfig{file=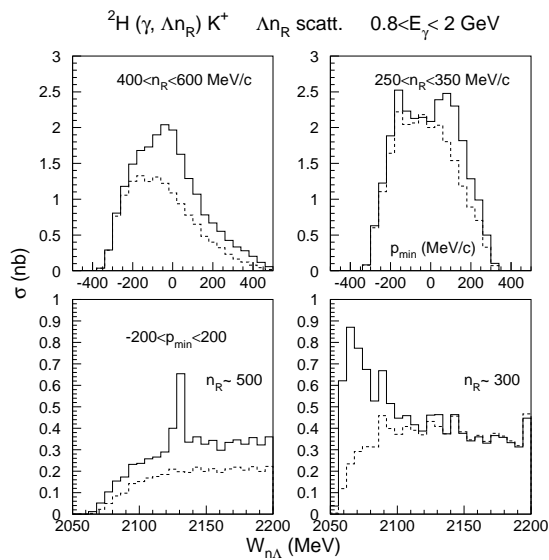,width=3.0in}
\caption[]{The $\Lambda n$  rescattering sector in the $\gamma ^2$H$\rightarrow K^+ n\Lambda$ reaction. The dashed histograms include the quasi-free process only. The full histograms correspond to the full model. See text for the definition of the cuts.}
\label{nL}
\end{center}
\end{figure}

Fig.~\ref{nL} shows observables which emphasize the $\Lambda n$ rescattering sector in the $\gamma ^2$H$\rightarrow K^+ n\Lambda$ reaction. Now, the minimum momentum $p_{min}(n\Lambda)$ is the lowest value of the momentum of the spectator neutron for which the $\Lambda$ can propagate on-shell in the Lambda nucleon scattering loop.  In the bottom parts the cut $-200<p_{min}(n\Lambda)<200$~MeV/c emphasizes $\Lambda n$ rescattering. For the low recoil momentum ($\sim 300$~MeV/c) bands (bottom right), strong S-wave rescattering enhances the  distribution of the mass $W_{\Lambda n}$ of the $\Lambda n$ system near the $\Lambda$ production threshold, but the quasi free contribution is still dominant above. For the high recoil momentum ($\sim 500$~MeV/c) bands (bottom left), CLAS acceptance suppresses the kinematics region near the $\Lambda$ production threshold but the quasi free process contribute less above. The cusp in the $\Lambda n$ scattering amplitude clearly appears at the $\Sigma$ production threshold ($W_{n\Lambda}= 2129$~MeV). The width of each bin in the $\Lambda n$ mass distribution is $6$~MeV, comparable to the CLAS mass resolution.

I defer the evaluation of inelastic rescattering to a forthcoming paper. On the one hand, $\Sigma$ photoproduction followed by the $\Sigma N\rightarrow \Lambda N$ transition may contribute below the $\Sigma$ production threshold. However, in the Regge model the $\Sigma$ photoproduction amplitude is much smaller than the $\Lambda$ one and the corresponding amplitude is not expected to dominate, at least at high energy. On the other hand, pion photoproduction followed by the $\pi N\rightarrow K^+ \Lambda$ may be significant~\cite{Sal04} but contributes in a very specific part of the phase space that emphasizes on shell $\pi N$ scattering. This contribution can be eliminated by a cut in a bi-dimensional plot similar to Fig~7 in ref.~\cite{La05}.

The signals are the same in the $\gamma ^2$H$ \rightarrow p K^{\circ} \Lambda$ reaction. In the quasi-free and $\Lambda N$ rescattering amplitudes, the elementary $K^{\circ}$ photoproduction amplitude~(\ref{elem_amp}) has to be used instead of the $K^+$ production one. In the $KN$ rescattering amplitude the combination of the elementary matrix elements in eq.~(\ref{pi_rescat}) becomes:
\begin{eqnarray}
\begin{array}{rcr}
T_{\gamma N}T_{K N}&=& -T^{K}_{\gamma p\rightarrow K^+ \Lambda}
\left( 0.5 \;T^1_{K N} -  0.5\; T^0_{K N}\right) \\
&&\\
&-& T^{K^*}_{\gamma p\rightarrow K^+ \Lambda}
\left( 1.27 \;T^1_{K N} +  0.271 \;T^0_{K N}\right)
\end{array}
\end{eqnarray}

In order to extract the final signals from experiment the model remains to be folded with the actual acceptance and efficiency of CLAS and the statistics must be significantly increased in the Monte Carlo sampling. However, the crude modeling of the acceptance and resolution that was used to get Figs.~\ref{Kn} and~\ref{nL} is good enough to tell us what would be the size of the signal of a possible pentaquark in the $KN$ scattering or a cusp in the $\Lambda N$ scattering sectors:  their ratio to the side band contribution is already on solid ground.

In very well defined part of the phase space, $K N$ and $\Lambda N$ rescattering depend on on-shell elementary matrix elements and on the low momentum components of the deuteron wave function. This provides us with the unique opportunity to study details in the  scattering amplitudes that may have escaped  the analysis of reactions induced on a nucleon target by low quality Kaon and Hyperon beams at low energies. A narrow state with a width of 1~MeV would contribute by no more than 10\% to the $KN$ mass spectrum, in accord with the lack of signal in the analysis~\cite{penta} of the most recent high statistic experiment~\cite{g10}.  On the contrary, this experiment allows a precise determination of the $\Lambda N$ scattering amplitude which would provide stringent constraints on Hyperon-Nucleon potential. In particular, a cusp would be easily detected near the $\Sigma$ production threshold in the $\Lambda N$ mass spectrum.

I acknowledge the warm hospitality at JLab where this work was completed. The Southern Universities Research Association (SURA) operates Thomas Jefferson National Accelerator Facility (JLab) for the US Department of Energy under Contract No DE-AC05-84ER40150.


\begin{thebibliography}{99}

\bibitem{g10}  K. Hicks and S. Stepanyan spokespersons, JLab Experiment E03-113.

\bibitem{La81} J.-M. Laget, Phys. Rep. {\bf 69}, 1 (1981).

\bibitem{La05} J.-M. Laget, Phys. Rev. C {\bf 73}, 044003 (2006); nucl-th/0507035.

\bibitem{PaXX} M. Lacombe {\it et al.}, Phys. Lett. {\bf B101}, 139 (1981).

\bibitem{La97} M. Guidal, J.M. Laget and M. Vanderhaeghen, Phys. Lett.  {\bf B400}, 6 (1997); Nucl. Phys. {\bf A627}, 645 (1997); Phys. Rev. C {\bf 57}, 1454 (1998).

\bibitem{Sch05} R. Bradford {\it  et al.} Phys. Rev. C {\bf 73} 035202 (2006); nucl-ex/0509033.

\bibitem{La05b} J.-M. Laget, Phys. Lett. {\bf B609}, 49 (2005).

\bibitem{PDG} S. Eidelman {\it et al.}, Phys. Lett. {\bf B572}, 1 (2004).

\bibitem{La93} J.-M. Laget, Journal of the Korean  Physical Society {\bf 26}, S244 (1993).

\bibitem{Gi74} G. Giacometti {\it et al.}, Nucl. Phys. {\bf B20}, 301 (1970); Nucl. Phys. {\bf B71}, 138 (1974).

\bibitem{La91} J.-M. Laget, Phys.Lett. {\bf B259}, 24 (1991).

\bibitem{Bo70} T. Bowen {\it et al.}, Phys. Rev. D {\bf 2}, 2599 (1970).

\bibitem{LEPS} T. Nakano {\it et al.}, Phys. Rev. Lett. {\bf 91}, 012002 (2003).

\bibitem{Ste03} S. Stepanyan {\it et al.}, Phys. Rev. Lett. {\bf 91}, 252001 (2003).

\bibitem{Kw05} H. Kwee, M. Guidal, M. Polyakov and M. Vanderhaeghen, Phys. Rev. D {\bf 72}, 054012 (2005).

\bibitem{Ha03} J. Haidenbauer and G. Krein, Phys. Rev. C {\bf 68}, 052201(R) (2003).

\bibitem{Ca04} R.N. Cahn and G.H. Trilling, Phys. Rev. D {\bf 69}, 011501(R) (2004).

\bibitem{Nu03} S. Nussinov, hep-ph/0307357

\bibitem{Ar03} R.A. Arndt, I.I. Strakosky and R.L. Workman, Phys. Rev. C {\bf 68}, 042201(R) (2003).

\bibitem{CLAS} B.A. Mecking {\it et al.}, Nucl. Instr. Meth. A {\bf 503}, 513 (2003).

\bibitem{La06} J.-M. Laget, nucl-th/0603009 v1.

\bibitem{Gu04} V. Guzey, Phys. Rev. C {\bf 69}, 065203 (2004).

\bibitem{Na79} M.M. Nagels, T.A. Rijken and J.J. de Swart, Phys. Rev. D {\bf 15}, 2547 (1977);  Phys. Rev. D {\bf 20}, 1633 (1979).

\bibitem{Rij99} T.A. Rijken {\it et al.}, Phys. Rev. C {\bf 59}, 21 (1999).

\bibitem{Sal04} A. Salam and H. Arenh\"{o}vel, Phys. Rev. C {\bf 70}, 044008 (2004).

\bibitem{penta} S. Niccolai {\it et al.}, hep-ex/0604047.

\end{thebibliography}
\end{document}